\documentclass[superscriptaddress, aps, showpacs, floatfix, prb,twocolumn]{revtex4-2}  
\usepackage{graphicx}% Include figure files
\usepackage{dcolumn}% Align table columns on decimal points
\usepackage{bm}% bold math
\usepackage{xcolor}
\usepackage{hyperref}% add hypertext capabilities

\usepackage{amsmath}

\def\BFO/{BiFeO$_3$}
\newcommand{\wn}{\,cm$^{-1}$\,}

\begin{document}

\title{THz spin-wave excitations in the transverse conical phase of \BFO/}

\author{B. T{\'o}th}
\email{toth.boglarka@ttk.bme.hu}
\affiliation{Department of Physics, Institute of Physics, Budapest University of Technology and Economics, M\H{u}egyetem rkp. 3. H-1111, Budapest, Hungary}

\author{D. G. Farkas}
\affiliation{Department of Physics, Institute of Physics, Budapest University of Technology and Economics, M\H{u}egyetem rkp. 3. H-1111, Budapest, Hungary}

%\author{I. K\'ezsm\'arki}
%\affiliation{Experimental Physics V, Center for Electronic Correlations and Magnetism, Institute of Physics, University of Augsburg, 86159 Augsburg, Germany}

\author{K. Amelin}
\author{T. R{\~o}{\~o}m}
\author{U. Nagel}
\affiliation{National Institute of Chemical Physics and Biophysics, Akadeemia tee 23, 12618 Tallinn, Estonia}

\author{L. Udvardi} 
\affiliation{Department of Theoretical Physics, Institute of Physics, Budapest University of Technology and Economics, M\H{u}egyetem rkp. 3. H-1111, Budapest, Hungary}

\author{L. Szunyogh}
\affiliation{Department of Theoretical Physics, Institute of Physics, Budapest University of Technology and Economics, M\H{u}egyetem rkp. 3. H-1111, Budapest, Hungary}
\affiliation{HUN-REN–BME Condensed Matter Physics Research Group, Budapest University of Technology and Economics, M\H{u}egyetem rkp. 3., H-1111 Budapest, Hungary}

\author{L. R{\'o}zsa}
\affiliation{Department of Theoretical Solid State Physics, Institute for Solid State Physics and Optics, HUN-REN Wigner Research Centre for Physics, H-1525 Budapest, Hungary}
\affiliation{Department of Theoretical Physics, Institute of Physics, Budapest University of Technology and Economics, M\H{u}egyetem rkp. 3. H-1111, Budapest, Hungary}

\author{T. Ito}
\affiliation{National Institute of Advanced Industrial Science and Technology (AIST), Tsukuba, Ibaraki 305-8562, Japan}

\author{S. Bord{\'a}cs}
\affiliation{Department of Physics, Institute of Physics, Budapest University of Technology and Economics, M\H{u}egyetem rkp. 3. H-1111, Budapest, Hungary}
\affiliation{HUN-REN–BME Condensed Matter Physics Research Group, Budapest University of Technology and Economics, M\H{u}egyetem rkp. 3., H-1111 Budapest, Hungary}
\affiliation{Experimental Physics V, Center for Electronic Correlations and Magnetism, University of Augsburg, D-86135 Augsburg, Germany}

\begin{abstract}
Although \BFO/ is one of the most studied multiferroic materials, recent magnetization and neutron scattering studies have revealed a new magnetic phase in this compound\textemdash the transverse conical phase. To study the collective spin excitations of this phase, we performed THz spectroscopy in magnetic fields up to 17\,T at and above room temperature. We observed five spin-wave branches in the magnetic phase with long wavelength conical modulation. Using a numerical spin dynamics model we found two kinds of excitations with magnetic moments oscillating either along or perpendicular to the static fields. Remarkably, we detected strong directional dichroism, an optical manifestation of the magnetoelectric effect, for two spin-wave modes of the conical phase. According to our experiments, the stability of the conical state is sensitive to the magnetic field history and it can become (meta)stable at or close to zero magnetic field, which may allow exploiting its magnetoelectric properties at room temperature.
\end{abstract}

\maketitle

\section{Introduction}

Among magnetoelectric multiferroics, i.e.,~materials with coexisting ferroelectric and magnetic orders \cite{Schmid1994, Hill2000, Kimura2003, Fiebig2019}, \BFO/ has been attracting a particular interest \cite{Park2014} as 1) it is one of the few room-temperature multiferroics \cite{Fiebig2019}, 2) shows a large ferroelectric polarization, even compared with conventional ferroelectrics \cite{Wang2003}, and 3) has a complex magnetic structure shaped by competing interactions \cite{Popov1994, Fishman2013}. Based on the magnetoelectric (ME) coupling between the electric polarization and the magnetic order, several potential applications of multiferroics were envisaged such as magnetoelectric memory and logic devices. \cite{Bibes2008, Chu2008, Heron2014, Manipatruni2019, Ramesh2019, Spaldin2019}. \BFO/ has been a test-bed for these concepts and many fundamental features relevant for applications, such as the electric field control of magnetic domains \cite{Zhao2006} and magnetisation \cite{Chu2008} or non-reciprocal light absorption \cite{Kezsmarki2015}, were demonstrated in this material.  

% magnonics--Kruglyak2010 

Even though \BFO/ has been studied extensively, large, ferroelectric monodomain single crystals became available only in the last ten years \cite{Ito2011}. These laser-floating-zone-grown crystals allowed the systematic characterization of the magnetic field-induced electric polarization \cite{Tokunaga2015, Kawachi2019}, the observation of the magnetic field-induced reorientation of the cycloidal domains \cite{Bordacs2018}, determining the optical selection rules of the spin-waves \cite{Farkas2021} and measuring their anisotropy \cite{Zhang2022}. Remarkably, the magnetic field-temperature phase diagram of \BFO/ is still not fully explored: recently a new magnetic phase, the transverse conical state was discovered above 150\,K \cite{Kawachi2017,Matsuda2020}. Although the spin-wave excitations of \BFO/ have been studied close to room temperature using INS \cite{Xu2012, Matsuda2012, Zhang2022}, Raman \cite{Rovillain2009} and THz \cite{Talbayev2011, Kezsmarki2015, Matsubara2016, Bialek2019, Liu2020}, the spin-wave excitation spectrum in the transverse conical phase remained unexplored. Here, we study the spin-wave excitations in this magnetic-field-induced transverse conical phase, but before presenting the results, we briefly summarize the most important aspects of the phase diagram of bulk \BFO/ below.

\begin{figure}[h]
\centering
\includegraphics[width = \linewidth]{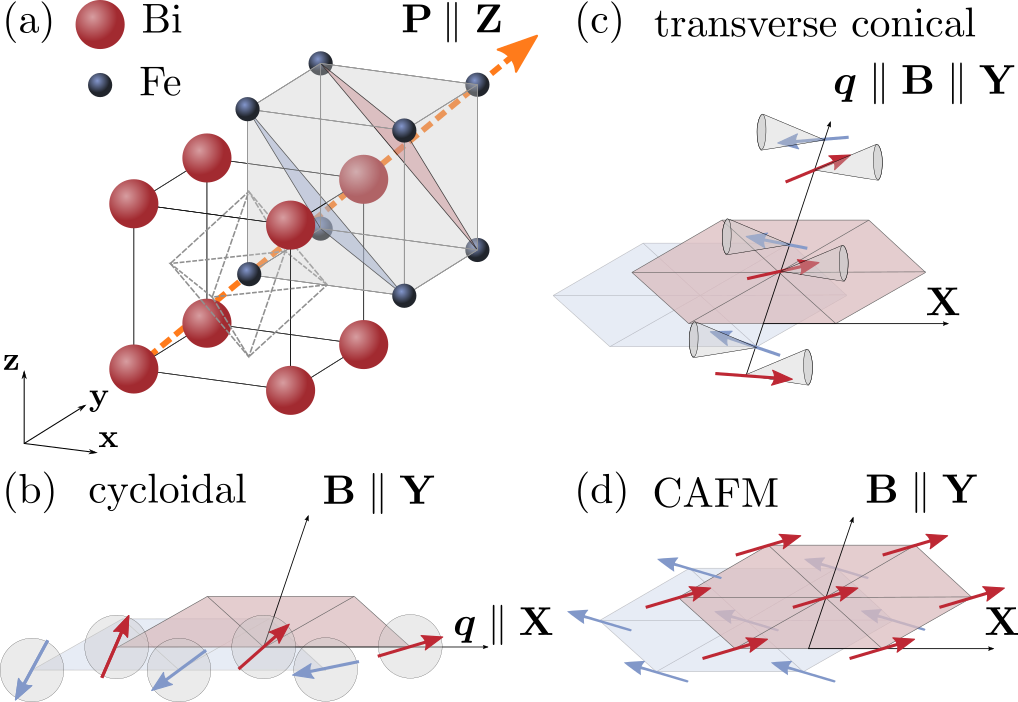}
%\captionsetup{justification=Justified}
\caption{(a) Perovskite unit cell of \BFO/. $\{\mathbf{x},\mathbf{y}, \mathbf{z}\}$ axes show the coordinate system of the pseudocubic unit cell. $\mathbf{Z}=[111]/\sqrt{3}$ axis is parallel to the hexagonal axis while $\mathbf{X}=[1\bar{1}0]/\sqrt{2}$ and $\mathbf{Y}=[11\bar{2}]/\sqrt{6}$ are in the perpendicular plane. At 1100\,K, the unit cell undergoes a rhombohedral distortion, and electric polarization emerges along a body diagonal. (b)--(d) Schematic drawings of magnetic structures in the different phases of \BFO/ when the static field $\mathbf{B}$ points along $\mathbf{Y}$. Spins in two antiferromagnetically coupled hexagonal planes, light red and light blue, are shown.}
\label{fig:structure}
\end{figure}

The perovskite structure of \BFO/ (see Fig.~\ref{fig:structure}(a)) undergoes a lattice distortion at around 1100\,K and its symmetry decreases to the rhombohedral $R3c$  \cite{Moreau1971,Kubel1990}. Upon this structural transition, ferroelectric polarization appears along one of the $\langle$111$\rangle$-type body diagonals \cite{Sosnowska1982}. Throughout the paper, we use the Cartesian system \{$\mathbf{X}$, $\mathbf{Y}$, $\mathbf{Z}$\} with axes parallel to $\{[1\bar{1}0], [11\bar{2}], [111]\}$, as shown in Fig.~\ref{fig:structure}(a). In the rhombohedral phase, below 640\,K, a cycloidally modulated antiferromagnetic order sets in (see Fig.~\ref{fig:structure}(b)) \cite{Kiselev1963}, thus, \BFO/ becomes multiferroic well above room temperature. The strongest Heisenberg interaction between the neighbouring spins is antiferromagnetic, favouring a G-type order \cite{Matsuda2012, Jeong2012}, i.e., with the magnetic moments pointing oppositely in consecutive atomic layers stacked along the [111] direction. The polar displacement of the ions and the antiferrodistortive rotation of the oxygen octahedra allow for a homogeneous and a staggered Dzyaloshinskii-Moriya (DM) interaction \cite{Ederer2005}. In competition with the Heisenberg interaction, the homogeneous DM interaction leads to the cycloidal modulation of the antiferromagnetic structure. The staggered component is responsible for a local weak-ferromagnetic component appearing perpendicular to the ferroelectric polarization. As the spins of the cycloidal structure rotate, this term results in an additional spin density wave (SDW) modulation \cite{Ramazanoglu2011}. The q-vectors describing the modulated order lie in the plane perpendicular to the ferroelectric polarization and parallel to one of the three equivalent $\langle$1$\bar{1}$0$\rangle$-type ($\mathbf{X}$-type) directions \cite{Sosnowska1982}.

In a magnetic field applied in the XY plane, the q-vectors rotate in the plane %to be 
perpendicular to the field and at low temperature, above 7\,T only domains with q-vectors (nearly) perpendicular to the field remain \cite{Bordacs2018, Farkas2021}. This rearrangement is mainly driven by the difference between the longitudinal and the transverse susceptibilities defined with respect to the local antiferromagnetic order parameter. A sufficiently high magnetic field suppresses the modulated order and a two-sublattice canted antiferromagnetic (CAFM) order with a finite weak magnetic moment along the applied field (see Fig.~\ref{fig:structure}(d) becomes energetically favourable \cite{Tokunaga2015, Room2020, Ruette2004, Popov1994}.
%As the net magnetization coming from the local weak-ferromagnetic component of the modulation, i.e., the SDW \cite{Ramazanoglu2011}, is zero for a cycloid having only even harmonics.

At temperatures above 150\,K, recent magnetization and polarization measurements \cite{Kawachi2017} and neutron scattering studies \cite{Matsuda2020} revealed an intermediate phase between the cycloidal and the CAFM phase. In agreement with former theoretical predictions \cite{Gareeva2013}, this transverse conical state with q-vectors parallel to the field emerges as a compromise: it is a combination of a spiralling and a homogeneous antiferromagnetic order as shown in Fig.~\ref{fig:structure}(c). Remarkably, this magnetic-field-induced phase possesses a linear magnetoelectric effect of 210\,ps/m \cite{Kawachi2017}, that is exceptionally large even compared to orthophosphates or boracites \cite{Astrov, Schmid1997, Rivera1994}.

Spectroscopic techniques have provided valuable information on the complex magnetic structure of \BFO/ and the interactions stabilizing it. Inelastic neutron scattering experiments revealed the magnon dispersion over the Brillouin zone, from which the nearest and next-nearest neighbor Heisenberg interactions were deduced \cite{Jeong2012,Matsuda2012}. Due to the long-wavelength modulation, the magnetic Brillouin zone is much smaller, which leads to a series of excitations close to the center of the Brillouin zone \cite{deSousa2008,Fishman2013}. These spin-wave excitations of the cycloidal phase were probed by high-resolution Raman and THz spectroscopy \cite{Cazayous2008,Talbayev2011,Nagel2013,Kezsmarki2015,Farkas2021}. The strength of the DM interactions and the easy-axis anisotropy were determined from the resonance frequencies. Moreover, the dynamic magnetoelectric coupling gives rise to non-reciprocal directional dichroism (NDD) in \BFO/, which is the light absorption difference for counter-propagating beams \cite{Kezsmarki2015}. However, despite these efforts the collective excitations of the transverse conical phase have not been studied so far.

In this paper, we present the magnetic field dependence of the THz spectra measured in ferroelectric monodomain \BFO/ samples at 300\,K and 350\,K up to 17\,T. At such high temperatures, our magnetic field range covers all the above-mentioned three phases. Besides the resonances of the cycloidal and the CAFM phases, we observed five THz active resonances in the transverse conical phase. In increasing fields, three resonances disappear and the remaining two modes continuously evolve to the resonances of the CAFM state. Remarkably, we observed strong NDD for two modes of the conical phase. Furthermore, we derived spin model parameters 
which describe well the resonances of the cycloidal and
CAFM phases, and qualitatively capture the spin-wave
modes of the conical phase near room temperature.

%Furthermore, we applied spin dynamics simulations, which describe well the resonances of the cycloidal and CAFM phases, and qualitatively capture the spin-wave modes of the conical phase near room temperature.

\section{Experimental Methods}
Fourier-transform infrared spectroscopy was performed on high-symmetry cuts of ferroelectric monodomain single crystals of \BFO/. Three pieces, large face normal to $\mathbf{X}$,$\mathbf{Y}$ and $\mathbf{Z}$ directions, were the same as used in works \cite{Room2020, Farkas2021}. Samples were large enough to cover a 3\,mm diameter hole; thickness of the samples was about 0.5\,mm and the samples were polished to a wedged shape with the 2 degree angle. Same experimental set-up as in Ref.~\cite{Farkas2021} was used to study THz absorption in magnetic field  but the Faraday and Voigt probes were modified. Sample was placed on a small copper disk with the heater and thermometer. The sample disk was thermally isolated from the rest of the probe using plastic parts. On the optical path between the sample and the 0.3\,K bolometer, four radiation shields with 4\,mm diameter aperture were installed. Despite these measures taken, the thermal heat load from the sample probe at 300\,K on the bolometer using the 40\,\wn high-cutoff filter reduced the bolometer sensitivity about one order of magnitude as compared to sample temperature of 4\,K.

Magnetic-field dependence of the THz transmission was measured up to 17 \,T in two configurations. In Faraday configuration, the light propagation vector $\mathbf{k}$ is parallel to the external magnetic field $\mathbf{B}$, whereas in Voigt geometry, $\mathbf{k}$ is perpendicular to $\mathbf{B}$. Before recording the spectra at 300 and 350\,K, a magnetic field of 17\,T was applied at low temperatures. This created a magnetic monodomain with the cycloidal-order q-vectors aligned perpendicular to the applied field. We obtained the absorption spectrum following Ref.~\cite{Farkas2021}. First, we determined the field-induced absorption difference $\Delta\alpha_B$ as %The absorption spectrum was measured without a reference hole, tracking only the field-induced absorption difference $\Delta\alpha$ as
\begin{equation}
    \Delta\alpha_B=\alpha(\mathit{B})-\alpha(0\,\mathrm{T}) = -\frac{1}{d}\mathrm{ln}\left( \frac{\mathcal{I}(\mathit{B})}{\mathcal{I}(0\,\mathrm{T})} \right),
\end{equation}
where $\mathcal{I}(\mathit{B})$ is the intensity of the transmitted light at magnetic field $B$, and $d$ is the sample thickness. Next, the zero-field spectrum is recovered by calculating the median of the absorption coefficient at each frequency, which reproduces the absorption at the resonances as they are well separated and shift considerably with the field. Finally, the zero-field spectrum is added to $\Delta\alpha_B$ to calculate $\Delta\alpha$. We obtained directional dichroism as the absorption difference detected upon the reversal of the magnetic field instead of the light propagation vector following earlier experiments \cite{Kezsmarki2015}.

%where $\mathcal{I}(\mathit{B})$ is the intensity of the transmitted light at magnetic field $B$, and $d$ is the sample thickness. As a possible systematic error, we mention that diffraction on the sample with a size comparable to the wavelength may result in a frequency-dependent baseline to the absorption spectra. Finally, we observed directional dichroism as the absorption difference detected upon the reversal of the magnetic field instead of the light propagation vector following earlier experiments \cite{Kezsmarki2015}.

\section{Experimental Results}

\subsection{Magnetic-field dependence of the spin-waves}

\begin{figure}[h]
\centering
\includegraphics[width = \linewidth]{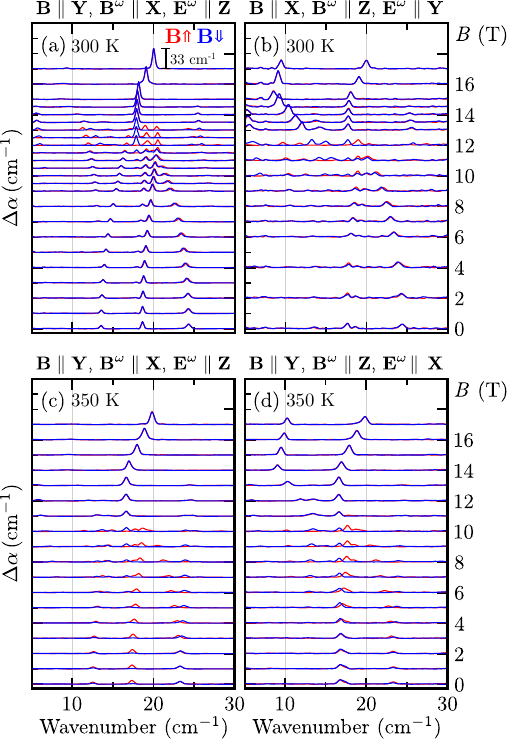}
%\captionsetup{justification=Justified}
\caption{ (a)--(d) Magnetic field dependence of the differential absorption spectra for various field directions and light polarization. The spectra are shifted by a constant proportional to the magnetic field strength. Spectra measured in increasing/decreasing the magnetic fields are red/blue. Spectra in (a) and (b) were measured at $T$\,=\,300\,K, whereas, spectra in (c) and (d) were measured at $T$\,=\,350\,K.
%the cycloidal-conical transition occurs at 13\,T in increasing fields, in decreasing fields the conical state changes back to cycloidal at 12\,T. Spectra in (c) and (d) were measured at $T$\,=\,350\,K, and the cycloidal-conical transition is only visible in increasing fields at 11\,T. 
}
\label{fig:4panel}
\end{figure}

Absorption spectra measured in Faraday configuration at 300\,K and 350\,K are shown in Fig.~\ref{fig:4panel}. Spectra recorded in increasing/decreasing fields are plotted in red/blue. The zero-field spin-wave resonances of the cycloidal state shift to lower frequencies as the temperature is increased, in agreement with former Raman and THz spectroscopy studies \cite{Cazayous2008,Talbayev2011,Bialek2019}, and follow a similar field dependence observed at low temperatures \cite{Nagel2013,Kezsmarki2015,Farkas2021}. However, in higher fields, there are two pronounced anomalies in the character of the spectra corresponding to the two critical fields, $B_{c1}$ and $B_{c2}$. On the border between the cycloidal and the transverse conical phases, at $B_{c1}$ the resonance frequencies suddenly change, whereas the appearance of the CAFM phase at $B_{c2}$ is indicated by the reduction of the number of modes to two and by kinks in the magnetic field dependence of the frequencies.

%\begin{figure*}[h]
%\centering
%\includegraphics[width = \linewidth]{fig1.png}
%\captionsetup{justification=Justified}
%\caption{a) Perovskite unit cell of \BFO/. At 1100\,K, the unit cell undergoes a distortion along the body diagonal, and electric polarization emerges along the distortion axis. b)--d) Schematic drawings of magnetic structures in the different phases of \BFO/. e)--h) Differential absorption spectra with various external field and light polarization configurations. Plotted spectra were added a constant to shift proportional to the external magnetic field. To capture the hysteresis of the conical phase, increasing and decreasing field spectra were both plotted. Spectra measured upon increasing the external fields are red, spectra recorded in decreasing fields are blue. Spectra in e), f) were measured at T\,=\,300\,K, the cycloidal-conical transition occurs at 13\,T in increasing fields, in decreasing fields the conical state changes back to cycloidal at 12\,T. Spectra in g), h) were measured at T\,=\,350\,K, and the cycloidal-conical transition is only visible in increasing fields at 11\,T. }
%\label{fig:fig1_general}
%\end{figure*}

Figure~\ref{fig:4panel} (a) and (b) present THz spectra recorded at $T$\,=\,300\,K in external fields $\mathbf{B}\parallel\mathbf{Y}$ and $\mathbf{B}\parallel\mathbf{X}$, respectively. %The polarization of the radiation was set parallel to $\mathbf{B}^{\omega}\parallel\mathbf{X}$ and $\mathbf{E}^{\omega}\parallel\mathbf{Z}$, and  $\mathbf{B}^{\omega}\parallel\mathbf{Z}$ and $\mathbf{E}^{\omega}\parallel\mathbf{Y}$, respectively.
When the field is increased along $\mathbf{B}\parallel\mathbf{Y}$, $B_{c1}$ is found between 13\,T and 13.5\,T, and the CAFM phase is reached at $B_{c2}$=15\,T. The two strongest modes of the conical state continuously transform into the modes of the two-sublattice CAFM phase in accordance with the second-order nature of the transition. When decreasing the magnetic field, the CAFM-conical transition occurs at the same field, $B_{c2}$=15\,T, but the conical phase persists down to 12\,T. When $\mathbf{B}\parallel\mathbf{X}$, the transition from the cycloidal to the conical state at $B_{c1}$ takes place between 12\,T and 13\,T in increasing fields. The CAFM phase is reached at the same field, $B_{c2}$=15\,T, as for $\mathbf{B}\parallel\mathbf{Y}$. Upon decreasing the field, the conical state remains stable until 12\,T, then the magnetic state changes back to cycloidal. The difference in $B_{c1}$ for $\mathbf{B}\parallel\mathbf{X}$ and $\mathbf{B}\parallel\mathbf{Y}$ detected in increasing fields may come from the in-plane magnetic anisotropy \cite{Bordacs2018,Fishman2018,Farkas2021}. Since the cycloidal-to-conical transition at $B_{c1}$ requires the q-vector to change its direction from perpendicular to parallel with the external field, the conical state for $\mathbf{B}\parallel\mathbf{X}$ has the lowest energy due to the in-plane anisotropy.

THz spectra measured at $T$\,=\,350\,K and in field $\mathbf{B}\parallel\mathbf{Y}$ are shown in Fig.\,\ref{fig:4panel} (c) and (d), for light polarization $\mathbf{B}^{\omega}\parallel\mathbf{X}$ and $\mathbf{E}^{\omega}\parallel\mathbf{Z}$, and $\mathbf{B}^{\omega}\parallel\mathbf{Z}$ and $\mathbf{E}^{\omega}\parallel\mathbf{X}$, respectively. The critical fields at this higher temperature are lower, as expected from the phase diagram of Ref.~\cite{Kawachi2017}. The cycloidal-to-conical phase transition occurs between 10\,T and 11\,T in increasing fields whereas the conical-to-CAFM phase boundary is at $B_{c2}$=14\,T. In decreasing fields, there is no discontinuous change indicating the transition from the transverse conical to cycloidal phase, i.e.,~the conical modes persist down to the lowest fields. After the application of a field, the spin-wave frequencies are the same in zero-field, though the magnitude of the resonances has changed. We may assume that at 350\,K, the transverse conical state stays metastable after high-field treatment or the conical state is continuously deformed to the cycloidal state. The stabilization of the conical state in small or even zero fields is compelling as its large magnetoelectric effect becomes available in low fields.

\begin{figure}[h]
\centering
\includegraphics[width = \linewidth]{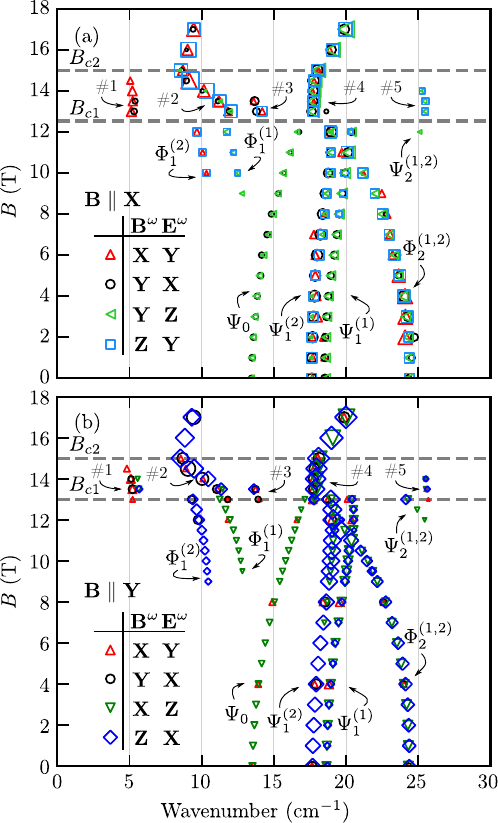}
%\captionsetup{justification=Justified}
\caption{Magnetic field dependence of spin-wave mode frequencies and intensities at 300\,K for magnetic fields $\mathbf{B}\parallel\mathbf{X}$, panel (a), and $\mathbf{B}\parallel\mathbf{Y}$, panel (b). Different symbols represent different light polarizations as given in the legends. The symbol size is proportional to the intensity, the absorption line area. Grey dashed horizontal lines mark the transition fields between the cycloidal-to-conical ($B_{c1}$) and the conical-to-CAFM ($B_{c2}$) phases. $\Phi$ and $\Psi$ respectively stand for the in-plane and out-of-plane modes of the cycloidal phase.
}
\label{fig:fig3_modes}
\end{figure}

To analyse the magnetic field dependence of the resonance frequencies deduced at $T$\,=\,300\,K, we plotted them in Fig~\ref{fig:fig3_modes} (a) and (b) for increasing fields applied along $\mathbf{X}$ and $\mathbf{Y}$, respectively. Each symbol refers to a specific polarization configuration and the symbol size is proportional to the mode strength. Horizontal dashed lines mark the critical fields $B_{\mathrm{c1}}$ and $B_{\mathrm{c2}}$. We note that the field steps are not equidistant in the different polarization configurations and different field regions. E.g. for $\mathbf{B}\parallel \mathbf{Y}$, $\mathbf{B}^{\omega}\parallel\mathbf{X}$ and $\mathbf{E}^{\omega}\parallel\mathbf{Z}$ (green down-pointing triangles in Fig.~\ref{fig:fig3_modes} (b)), spectra were recorded at every integer value of magnetic field between 0\,T and 9\,T, and every half \,T was measured between 9.5\,T and 15\,T, but for $\mathbf{B}\parallel \mathbf{Y}$, $\mathbf{B}^{\omega}\parallel\mathbf{X}$ and $\mathbf{E}^{\omega}\parallel\mathbf{Y}$ (red up-facing triangles in Fig.~\ref{fig:fig3_modes} (b)), only every fourth Tesla was measured between 0\,T and 12\,T, every half Tesla between 13\,T and 15\,T, and above that, only 17\,T.

\subsubsection{Cycloidal phase}
Altogether, we observed seven modes in the cycloidal phase that we labelled following Ref.~\cite{Farkas2021}. At the lowest energies, $\Phi_1^{(2)}$ gains intensity only in finite fields just like $\Phi_1^{(1)}$ at about 2\,\wn higher energy. $\Psi_0$, $\Psi_1^{(2)}$, $\Psi_1^{(1)}$ and $\Phi_2^{(1,2)}$ are present already in zero fields at 13.4\,\wn, 17.6\,\wn, 18.5\,\wn and 24.3\,\wn, respectively. When $\mathbf{B}\parallel\mathbf{Y}$, the weak $\Psi_2^{(1,2)}$ mode also gains strength in finite fields, appearing at 25.4\,\wn. The selection rules of these modes are mostly in agreement with the low-temperature results of Ref.~\cite{Farkas2021}. The only exception is  $\Phi_1^{(1)}$, that is observed in Ref.~\cite{Farkas2021} when $\mathbf{B}\parallel\mathbf{X}$, $\mathbf{B}^{\omega}\parallel\mathbf{Y}$ and $\mathbf{E}^{\omega}\parallel\mathbf{X}$, but is absent in our measurements. As this is a weak resonance gaining strength only in finite fields, it probably remains undetectable in the high-temperature experiments due to the reduced signal to noise ratio.

\subsubsection{Transverse conical phase}
In the transverse conical phase, between $B_{\mathrm{c1}}$ and $B_{\mathrm{c2}}$, we observed five modes, see Fig.~\ref{fig:4panel} and \ref{fig:fig3_modes}. The mode \#1 at 5.5\,\wn, \#2 starting at 12\,\wn and evolving into the lower-frequency mode of the CAFM phase. Mode \#3 begins at 14\,\wn, \#4, the nearly field-independent mode at 17.7\,\wn evolves to be the higher-frequency mode of the CAFM phase, and \#5 emerges at 25.5\,\wn. As the set of measured polarization configurations is incomplete, we can make only a few claims about the selection rules of these modes.
 
The two strongest modes of the conical phase, \#2 and \#4, evolve into the lower- and higher-energy modes of the CAFM phase. These resonances are present in all polarization configurations. However, mode \#2 is weak for $\mathbf{B}^{\omega}\parallel \mathbf{Y}$ when $\mathbf{B}\parallel \mathbf{X}$ and also for $\mathbf{B}^{\omega}\parallel \mathbf{X}$ when $\mathbf{B}\parallel \mathbf{Y}$ (see black and green symbols in Fig.~\ref{fig:fig3_modes} (a), and  red and green symbols in Fig.~\ref{fig:fig3_modes} (b)). In these configurations, its strength decreases with the field, while in all other configurations, it grows as the field increases. These observations suggest that the in-plane component of $\mathbf{B}^{\omega}$ perpendicular to the static field does not significantly couple to this resonance. Mode \#4 is weak for polarization $\mathbf{B}^{\omega}\parallel\mathbf{X}$ and $\mathbf{E}^{\omega}\parallel\mathbf{Y}$ for both external field directions (see red triangles in Fig.~\ref{fig:fig3_modes}). 

%\textcolor{blue}{
%Mode \#2 is one of the two modes in the conical phase, that evolves into the lower energy mode of the CAFM phase. When $\mathbf{B}\parallel \mathbf{X}$, it is present in all polarization configurations, but while for $\mathbf{B}^{\omega}\parallel \mathbf{Y}$, it is very weak and its strength decreases with the field, in all other configurations it is strong and grows stronger as the field increases. The same is true for $\mathbf{B}\parallel \mathbf{Y}$, but in that case the mode is weak and gets weaker when $\mathbf{B}^{\omega}\parallel \mathbf{X}$.
%}
%\textcolor{blue}{
%Mode \#4 is present in all configurations. It is weakly present and loses intensity for the polarization configuration of $\mathbf{B}^{\omega}\parallel\mathbf{X}$ and $\mathbf{E}^{\omega}\parallel\mathbf{Y}$ for both external field directions.
%}
%I do not see that the size of the red triangles changes.

The other three modes are weaker and lose their intensity with increasing fields. Mode \#1 with decreasing energy and strength is active in all configurations when $\mathbf{B}\parallel \mathbf{Y}$. Even though we observed this mode in only two polarization configurations for $\mathbf{B}\parallel \mathbf{X}$, we can not draw further conclusions as it is close to the low-frequency cut-off of the THz spectroscopy setup. Similarly, mode \#5 is close to the high-frequency cut-off set by the 40\wn low-pass  filter. For both $\mathbf{B}\parallel \mathbf{X}$ and $\mathbf{B}\parallel \mathbf{Y}$, mode \#3 is silent in one polarization configuration, for  $\mathbf{B}^{\omega}\parallel\mathbf{Y}$ and $\mathbf{E}^{\omega}\parallel\mathbf{Z}$, and for  $\mathbf{B}^{\omega}\parallel\mathbf{X}$ and $\mathbf{E}^{\omega}\parallel\mathbf{Z}$, respectively (green symbols in Fig.~\ref{fig:fig3_modes}). As neither of these fields can excite this mode, it is electric dipole active only for electric fields oscillating in-plane, while $\mathbf{B}^{\omega}$ along $\mathbf{Z}$ or $\mathbf{B}$ may also excite this resonance. 

%Since its frequency is close to the low-frequency cutoff, we cannot deny the same to be true for $\mathbf{B}\parallel \mathbf{X}$, even though there are only two polarization configurations when the mode is observed. 
%The energy and the {\red{strength}} of mode \#1 decreases with the field.

%\textcolor{blue}{
%For both $\mathbf{B}\parallel \mathbf{X}$ and $\mathbf{B}\parallel \mathbf{Y}$, mode \#3 is silent in one polarization configuration, for  $\mathbf{B}^{\omega}\parallel\mathbf{Y}$ and $\mathbf{E}^{\omega}\parallel\mathbf{Z}$, and for  $\mathbf{B}^{\omega}\parallel\mathbf{X}$ and $\mathbf{E}^{\omega}\parallel\mathbf{Z}$, respectively (green symbols in Fig.~\ref{fig:fig3_modes}). This can be interpreted as the electric dipole can be excited with in-plane oscillating fields, while the magnetic dipole can be excited by oscillating magnetic fields perpendicular to the XY-plane, but containing the axis of the external magnetic field, that is also the direction of the $q$-vector of the conical structure. For both external field directions this mode becomes weaker as the magnetic field increases, and vanishes well before $B_{c2}$ is reached.
%} 
%\textcolor{blue}{
%Mode \#5, as mode \#3 and \#1,  also becomes weaker with increasing field. It is also near the high-frequency noise limit of some measurements, therefore selection rules are not possible to deduce.
%}

\subsubsection{CAFM phase}
In the high-field CAFM phase, two resonances emerge from the strongest modes of the conical phase. 
%The lower-frequency mode evolves from mode \#2 of the conical phase, and goes up in energy with the external field. 
The lower-frequency mode is strong in two polarization configurations for both external field directions. When  $\mathbf{B}\parallel \mathbf{X}$, those are  $\mathbf{B}^{\omega}\parallel\mathbf{X}$ and $\mathbf{E}^{\omega}\parallel\mathbf{Y}$ and  $\mathbf{B}^{\omega}\parallel\mathbf{Z}$ and $\mathbf{E}^{\omega}\parallel\mathbf{Y}$. And when  $\mathbf{B}\parallel \mathbf{Y}$, the favored polarizations are  $\mathbf{B}^{\omega}\parallel\mathbf{Y}$ and $\mathbf{E}^{\omega}\parallel\mathbf{X}$ and  $\mathbf{B}^{\omega}\parallel\mathbf{Z}$ and $\mathbf{E}^{\omega}\parallel\mathbf{X}$. This lower-frequency mode is also present for $\mathbf{B}\parallel \mathbf{X}$, $\mathbf{B}^{\omega}\parallel\mathbf{Y}$ and $\mathbf{E}^{\omega}\parallel\mathbf{X}$, but much weaker than in the former cases. Therefore, the lower-frequency mode interacts weakly with in-plane $\mathbf{B}^{\omega}$ perpendicular to $\mathbf{B}$. The higher-frequency mode is present in all configurations, but very weak when $\mathbf{B}^{\omega}\parallel\mathbf{X}$ and $\mathbf{E}^{\omega}\parallel\mathbf{Y}$, for both $\mathbf{B}\parallel \mathbf{X}$ and $\mathbf{B}\parallel \mathbf{Y}$. As can be seen from Fig.~\ref{fig:fig3_modes} (a) and (b), the CAFM spin-wave frequencies do not depend on the magnetic-field direction in the hexagonal plane: they are the same for $\mathbf{B}\parallel\mathbf{X}$ and $\mathbf{B}\parallel\mathbf{Y}$. This is different than at liquid He temperatures, where the higher-energy mode shifted to higher energy when the magnetic field direction changed from $\mathbf{B}\parallel\mathbf{X}$  to $\mathbf{B}\parallel\mathbf{Y}$ \cite{Room2020}.

%Unlike in low temperatures~\cite{Room2020}, the resonance frequencies in the CAFM state do not depend on the orientation of a field perpendicular to $\mathbf{Z}$.

\subsection{Directional dichroism of the spin-waves}

\begin{figure}[h]
\centering
\includegraphics[width = \linewidth]{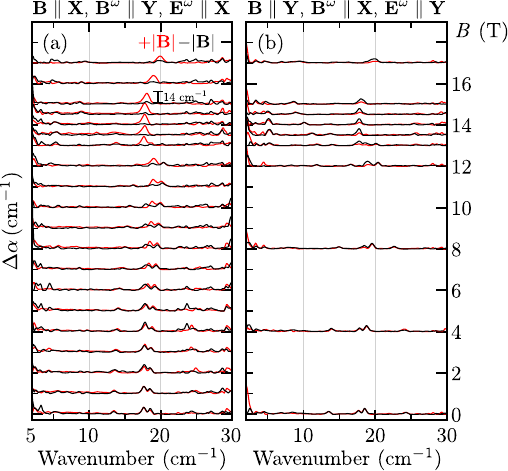}
%\captionsetup{justification=Justified}
\caption{Spectra measured at $T$\,=\,300\,K in positive/negative fields (red/black) with increasing magnitude are shown for various light polarizations. The difference between the absorption measured in positive and negative fields is the non-reciprocal directional dichroism (NDD).}
\label{fig:fig4_DD}
\end{figure}

A finite-frequency manifestation of the ME effect is the so-called NDD \cite{Tokura2007,Kezsmarki2011,Bordacs2012,Kezsmarki2015}, namely, that the light-absorption coefficient is different for light passing through a medium in opposite directions. In \BFO/, the reversal of the external magnetic field is considered to be equivalent to the reversal of the light-propagation direction \cite{Kezsmarki2015}, thus, we detected NDD as the difference of the absorption spectrum for positive and negative field directions. We measured absorption spectra for $\mathbf{k}\parallel\mathbf{Z}$ in two Voigt configurations, $\mathbf{B}\parallel\mathbf{X}$ and $\mathbf{B}\parallel\mathbf{Y}$. The results are presented in Fig.~\ref{fig:fig4_DD} for increasing field magnitudes at $T$\,=\,300\,K. The finite difference between the spectra measured in positive and negative fields is the NDD (compare red and black curves in Fig.~\ref{fig:fig4_DD}). The dichroism is present in all three phases and its magnitude in the THz range peaks at around the upper critical field, $B_{c2}\approx$15\,T.

Studying the sum rule of NDD spectra may provide information about the static ME susceptibility \cite{Szaller2014}:
\begin{equation}
    \chi_{\gamma \delta}^{\text{me}}(0)=\frac{c}{2\pi}\mathcal{P}\int_{0}^{\infty} \frac{\Delta \alpha (\omega)}{\omega ^2} d\omega,
\label{eq:sumrule}
\end{equation}
where $\Delta\alpha_{k} = \alpha_{+k}(\omega) - \alpha_{-k}(\omega)$ is the difference of the absorption spectra. The $\omega^{2}$ dependence of Eq.~(\ref{eq:sumrule}) implies that the strongest contribution the the static ME susceptibility comes from NDD of low frequency modes. In the present experimental configuration, when $\mathbf{B}\parallel\mathbf{X}$ and $\mathbf{B}^{\omega}\parallel\mathbf{Y}$ and $\mathbf{E}^{\omega}\parallel\mathbf{X}$, the contribution of the spin-waves to the ME susceptibility at 14\,T is 3.3\,ps/m as calculated with Eq.~(\ref{eq:sumrule}). 

The static ME effect of the conical phase has been studied in Ref.~\cite{Kawachi2017}, where they probed the longitudinal response, $\chi^{ME}_{YY}$ i.e.,~measured the magnetic-field-induced electric polarization in increasing magnetic fields. However, this experiment provides a different component of the ME tensor compared to our THz results. Here, the oscillating magnetic field $\mathbf{B}^{\omega}$ is perpendicular to the static field $\mathbf{B}$, i.e.,~it corresponds to the transverse response $\chi^{ME}_{XY}$ or $\chi^{ME}_{YX}$, which can be measured by either rotating a static field or by adding a small oscillatory field perpendicular to the static field. Thus, our results give an estimate of the spin-wave contribution to the static ME susceptibility, which may be verified in one of these configurations.

\section{Theoretical Modeling of Spin-wave resonances}

Following the literature \cite{Fishman2013}, we used the Hamiltonian below to calculate the spin-wave resonances of the various phases:
\begin{align}
    \mathcal{H} = &J_1 \sum_{\langle i,j\rangle} \mathbf{S}_i \cdot \mathbf{S}_j +J_2 \sum_{\langle i,j\rangle '} \mathbf{S}_i \cdot \mathbf{S}_j  \nonumber \\ 
    &+ D_1 \sum_{\langle i,j\rangle} \left(\mathbf{Z} \times \mathbf{e}_{ij}/a\right) \left(\mathbf{S}_i \times \mathbf{S}_j \right) \nonumber \\
    &+ D_2 \sum_{\langle i,j\rangle} \left(-1 \right)^{h_i}\mathbf{Z}\left(\mathbf{S}_i \times \mathbf{S}_j\right) \nonumber \\
    &- K \sum_i S_{iZ}^2 + g\mu_{\textrm{B}}\sum_i \mathbf{S}_i \cdot \mathbf{B}, \label{eq:Hamilton}
\end{align}
where $a$ is the pseudo-cubic lattice constant, see Fig.~\ref{fig:structure}. Each site, $\mathbf{R}_i$ is occupied by an Fe atom with a classical spin vector $\mathbf{S}_i$ of length S=5/2, and the nearest neighbours, $\langle i,j\rangle$ are connected by the vector $\mathbf{R}_j$=$\mathbf{R}_i$+$\mathbf{e}_{ij}$. $J_1$ and $J_2$ are the exchange coupling constants between nearest and next-nearest neighbours. $D_1$ and $D_2$ are the homogeneous and staggered components of the nearest-neighbour DM interaction, respectively, and $h_i$=$\sqrt{3}\mathbf{R}_i\cdot\mathbf{Z}$/a labels the hexagonal layers. The coupling constant of the easy-axis anisotropy is denoted by $K$. The g-factor is $g$=2, $\mu_{\textrm{B}}$ is the Bohr magneton and $\mathbf{B}$ is the magnetic field. The model parameters are collected in Table~\ref{tab:parameters}.

We performed the simulations for a cell consisting of 6 layers along the $\mathbf{Z}$ direction, enabling the use of periodic boundary conditions when considering the three alternating in-plane atomic positions in the pseudo-cubic lattice along this direction and the antiferromagnetic alignment between neighbouring layers. The system size was larger along the wave-vector direction, i.e., $\mathbf{X}$ perpendicular to the field direction for modelling the cycloidal structure and $\mathbf{Y}$ along the field for the conical spiral, while the size was kept minimal along the direction perpendicular to the long side. Periodic boundary conditions were used along all directions and a single period of the spin spiral was included in the cell. The period of the spiral, i.e., the length of the long side of the cell, was optimized at each magnetic field value by calculating the ground-state energy per spin for various cell sizes.

The energy was found by initializing a system in a harmonic spin cycloid. The energy of this state was minimized first by numerically solving the Landau–Lifshitz–Gilbert equation using the full Hamiltonian, then speeding up the relaxation close to the energy minimum by aligning the spins with the direction of the local effective magnetic field in each step. The convergence was stopped when the torque expressed in frequency units became smaller than $10^{-8}$\,meV$g/\hbar$ at each lattice site. The spin-wave frequencies were calculated within linear spin-wave theory close to this energy minimum. The spin-wave eigenvectors of each mode $n$ were used to determine the response function $\left|\chi_{0}^{n}\right|^{2}$ to a spatially homogeneous external magnetic field at the spin-wave frequency, i.e., the response function is proportional to the magnetic dipole strength of the mode. Details of the calculation method and the definition of the response function are reported in Ref.~\cite{Rozsa_2020}.

\begin{figure}[h]
\centering
\includegraphics[width = \linewidth]{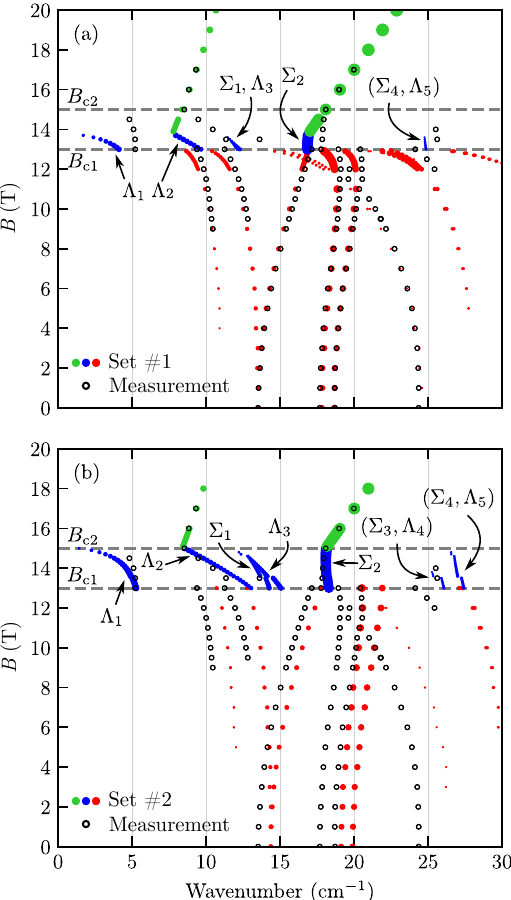}
%\captionsetup{justification=Justified}
\caption{Panel (a) and (b) present the resonance positions calculated using the model Hamiltonian defined in Eq.~(\ref{eq:Hamilton}) with parameter set \#1 and set \#2, respectively, as given in Table~\ref{tab:parameters}. Red, blue and
green dots show the theoretical frequencies in the cycloidal, transverse conical and CAFM phases, respectively. The size of the dots is proportional to the magnetic dipole strength of the given mode. $\Lambda_n$ and $\Sigma_n$ correspond to modes with oscillating magnetic moments parallel or perpendicular to the static field, respectively. The resonance frequencies measured in experiments in external magnetic fields parallel to $\mathbf{Y}$ are shown in black. Grey dashed horizontal lines mark the transition fields between the cycloidal-conical ($B_{c1}$) and the conical-CAFM ($B_{c2}$) phases in the experiments.}
\label{fig:sim_vs_meas}
\end{figure}

 %Red, blue and green dots show the theoretical frequencies in the cycloidal, transverse conical and CAFM phases, respectively. The size of the dots is proportional to the magnetic dipole strength of the given mode. The experimentally detected positions are shown with black circles. The area of the symbols is proportional to the intensity. 

\begin{table}
    \centering
    \begin{tabular}{c|c|c|c|c|c}
         & $J_1$ (meV) & $J_2$ (meV) & $D_1$ ($\mu$eV) & $D_2$ ($\mu$eV) & $K$ ($\mu$eV)\\
         \hline
        set \#1 & 4.9 & 0 & 177  & 77.3  & 3.1 \\
%        set \#2 & 4.88 & -0.45  & 191   & 68.9 & 2.96 \\
        set \#2 & 4.9 & 0.17  & 177   & 77.3 & 3.1
    \end{tabular}
    \caption{Exchange, DM and anisotropy parameters used in the simulations.}
    \label{tab:parameters}
\end{table}

The calculated spin-wave resonances are compared with the experimental results in Fig.~\ref{fig:sim_vs_meas}.
In Fig.~\ref{fig:sim_vs_meas} (a), excitations are calculated with parameter set \#1 given in Table~\ref{tab:parameters}. We optimized these coupling constants to fit the cycloidal wavelength (62\,nm), the zero-field resonances of the cycloidal state and the field dependence of the modes in the CAFM phase. In this phase, $D_2$ and the product $J_1 K$ determine the resonances, $\omega_{\nu_{2}}$ and $\omega_{\nu_{1}}$ \cite{Room2020}, in agreement with Ref.~\cite{Pincus1960}: 
\begin{align}
\omega_{\nu_{2}} &= \gamma\sqrt{B^2+BB_D}\\
\omega_{\nu_{1}} &= \gamma\sqrt{BB_D-2B_EB_A+B_D^2},
\end{align}
where $\gamma$ is the gyromagnetic ratio, and we introduced the following effective fields: $B_D$=$\frac{z_1D_2S}{g\mu_B}$, $B_E$=$\frac{z_1J_1S}{g\mu_B}$, $B_A$=$\frac{2KS}{g\mu_B}$, where $z_1$=6 is the number of nearest neighbors. The in-plane, $\Phi_{n}$ and out-of-plane, $\Psi_{n}$ excitation frequencies of the zero-field cycloidal state may be well approximated by $\omega_{\Phi_{n}}\approx\gamma \sqrt{B_{E}B_L^{2}/(2B_S)n^2}$ and $\omega_{\Psi_{n}}\approx\gamma\sqrt{B_{E}B_L^{2}/(2B_S)\left(n^{2}+1\right)}$, respectively, with $B_L$=$\frac{2D_{1}S}{g\mu_B}$ and $B_S$=$\frac{\left(J_{1}-4J_{2}\right)S}{g\mu_B}$~\cite{deSousa2008,Fishman2013}. The ratio %$D_1$/$J_1$ 
$2\pi \left(2B_S\right)/B_L$ is the period of the cycloid in units of $a$ in $\mathbf{B} = 0$. The period and the frequencies are perturbed by the anisotropy in combination $B_E B_A$. In order to reduce the number of free parameters, first, we set $J_2$=0. The four parameters of set \#1 describe well the magnetic field dependence of the resonance frequencies in the cycloidal and CAFM phases, and also the lower critical field, $B_{c1}$. Besides the frequencies, the numerical model correctly describes the motion of spins in the cycloidal phase as the modes labelled as $\Psi$ and $\Phi$ indeed correspond to the out-of-plane and in-plane oscillations of the spins. The parameter values in set \#1 are similar to, although somewhat lower than, the ones applied to describe the low-temperature resonances\cite{Kezsmarki2015, Farkas2021, Room2020}. In these publications, the same model was used apart from a sign difference in the definition of the exchange terms in the Hamiltonian, Eq.~(\ref{eq:Hamilton}). The decrease of the interaction parameters accounts for the lowered excitation frequencies in the high-temperature experiments

In the transverse conical state, the model qualitatively describes the field dependence of the resonances and gives an almost field-independent q-vector in accordance with recent neutron-scattering experiments \cite{Matsuda2020}. The eigenmodes can be classified into two groups. They possess an oscillating magnetic moment either parallel or perpendicular to the static field, thus, we label them $\Lambda_n$ and $\Sigma_n$, respectively. $\Lambda_1$, which is the lowest-energy mode of the conical phase, i.e., mode \#1 in Fig.~\ref{fig:fig3_modes} (a), corresponds to the opening and closing of the cone angle. As expected from this motion, it softens and loses its strength toward $B_{c2}$. In agreement with the experimental finding that mode \#2 is not significantly coupled to an in-plane $\mathbf{B}^{\omega}$ perpendicular to $\mathbf{B}$, we assign it to $\Lambda_2$. As mode \#3 shows a similar selection rule, it is likely $\Lambda_3$, though it may overlap with $\Sigma_1$ appearing at almost the same energy in the numerical calculations. Although we could not deduce clear selection rules for mode \#4, it is clearly the $\Sigma_2$ mode based on the calculations. Mode \#5 could be one of the pairs $\left(\Sigma_{3},\Lambda_{4}\right)$ or $\left(\Sigma_{4},\Lambda_{5}\right)$ which are almost degenerate in the calculations. The $\left(\Sigma_{3},\Lambda_{4}\right)$ pair is not visible in Fig.~\ref{fig:sim_vs_meas}(a) because of its low intensity. However, experimentally we could not distinguish between modes $\Sigma_{3}, \Sigma_{4}, \Lambda_{4}$ and $\Lambda_{5}$. Experimentally some of the modes are active in polarizations where the magnetic dipole selection rules do not predict their appearance; these may be excited by the oscillating electric field through the strong magnetoelectric effect detected in the conical phase.

Despite the qualitative agreement between theory and experiment, calculations predict lower frequencies in the conical phase and also a lower transition field, $B_{c2}\approx$13.8\,T. The upper critical field may be found as the field below which the CAFM phase becomes unstable against periodic modulations, i.e.,~the energy of a finite-q spin-wave becomes negative \cite{Kulagin2011}. The equation derived can be formally expressed with the frequencies of the modes in the CAFM phase at $B_{\mathrm{c2}}$ and the zero field cycloidal frequency, $\Psi_0$:
\begin{align}
%\omega^{2}_{Hi}\left(B_{c2}\right)\omega^{2}_{Lo}\left(B_{c2}\right)-\left[2\omega^{2}\left(\Psi,0\right)-\frac{\omega^{2}_{Hi}\left(B_{c2}\right)+\omega^{2}_{Lo}\left(B_{c2}\right)}{2}\right]^{2}=0.\label{Bc2eq}
\omega_{\nu_{2}}\left(B_{c2}\right)+\omega_{\nu_{1}}\left(B_{c2}\right)=2\omega_{\Psi_{0}}.\label{Bc2eq}
\end{align}
Therefore, once the spin-wave frequencies in the zero-field cycloidal state and in the CAFM phase are fixed, $B_{c2}$ cannot be modified by tuning the parameters. By introducing a finite $J_2$ in the original parameter set (see set \#2 in Table~\ref{tab:parameters}) the stability range of the modulated phases extends, and the CAFM phase appears above $B_{c2}\approx$15\,T  as shown in Fig.~\ref{fig:sim_vs_meas}(b). However, the energy of the conical phase is lower than that of the cycloidal phase only in a narrow field range above $B_{c1}$=14.9\,T. Although the conical state is not a global energy minimum below $B_{c1}$, it remains a local minimum, allowing us to calculate excitations above this metastable state. In Fig.~\ref{fig:sim_vs_meas}(b), we show the spin-wave resonances of the conical state in the 13-15\,T field range. The correspondence between theory and experiment becomes much better in the conical phase as the modes are shifted to higher frequencies. However, increasing $J_2$ leads to smaller exchange stiffness, thus, longer cycloidal q-vector and correspondingly increased spin-wave frequencies in the cycloidal phase.

We could not find a parameter set that captures the critical fields and the resonance frequencies in all three phases. The sizeable magnetostriction observed across the magnetic phase transition \cite{Kawachi2017} may explain why a single field-independent parameter set cannot describe all the experimental results. At low temperatures, the introduction of additional anisotropy parameters induced by magnetostriction explained the anisotropy of the resonances detected in the CAFM phase \cite{Room2020}. However, at high temperatures the magnetic resonances of the CAFM phase are not sensitive to the in-plane rotation of the field as seen in Fig.~\ref{fig:fig3_modes}. Moreover, the introduction of such additional anisotropies can modify the frequencies in the CAFM phase, but Eq.~\eqref{Bc2eq} remains valid. This means that additional anisotropies cannot change the upper critical field $B_{c2}$ if the resonance frequencies are fixed. When fixing $\omega_{\nu_{2}},\omega_{\nu_{1}}$ and $B_{c2}$, we could only take into account the magnetostriction by considering a different value of $\omega^{2}_{\Psi_{0}}=\gamma^{2}B_{E}B_{L}^{2}/(2B_S)$ %$\omega_{\textrm{cyc},0}$ 
in the cycloidal state and the CAFM phase. This coefficient is determined by parameters influencing the spatial variation of the structure, such as the exchange $J_{2}$ selected in parameter set \#2 or DM interactions, but is not significantly influenced by local anisotropy terms. Since we lack sufficient data to determine how all of the model parameters change as a consequence of the magnetostriction, we propose a minimal model where considering two different values of $J_{2}$ is sufficient to quantitatively describe the spin-wave frequencies in all three phases and the two critical fields.

\section{Summary}

We studied the spin-wave excitations of the transverse conical state of \BFO/ by THz spectroscopy. We determined the magnetic-field dependence of spin-wave excitations in all three phases at room temperature and above: the cycloidal, the transverse conical and the canted antiferromagnetic (CAFM) phases. Moreover, we analysed selection rules for the linearly polarized THz radiation absorption spectra. In the transverse conical phase we observed five modes, among which two evolve into the resonances of the CAFM phase. As a manifestation of the magnetoelectric effect, we also observed non-reciprocal directional dichroism, which is the strongest in the conical state. We developed a numerical model that describes well the resonances of the cycloidal and CAFM phases. The agreement between theory and experiment is qualitative in the conical phase indicating the importance of magnetostriction. Based on this model, we classified the resonances of the conical state into two groups: active for magnetic moments oscillating either along or perpendicular to the static field. Finally, we found that the conical state, which shows strong linear magnetoelectric effect and directional dichroism, may become (meta)stable at or close to zero magnetic field. Given the strain-tunability of the magnetic phases in \BFO/ \cite{Sando2013,Sando2014,Hemme2023}, thin-film technology may allow the realisation of these peculiar effects at room temperature.

\begin{acknowledgments}
We thank R.~S.~Fishman for the discussions. The authors acknowledge the support of the bilateral program of the Estonian and Hungarian Academies of Sciences under the contract NKM 2018-47, NKM 2021-24 and NKM2022-27/2023. We acknowledge support by the Estonian Research Council Grants No. PRG736  and by European Regional Development Fund Project No. TK134. This research was supported by the Ministry of Culture and Innovation and the National Research, Development and Innovation Office within the Quantum Information National Laboratory of Hungary (Grant No. 2022-2.1.1-NL-2022-00004). This work was supported by the Hungarian National Research, Development and Innovation Office – NKFIH grants FK 135003, K 131938, K 142652, and FK 142601, as well as by the Hungarian
Academy of Sciences via a János Bolyai Research
Grant (Grant No. BO/00178/23/11).  
\end{acknowledgments}

%\bibliography{bibem.bib}

%apsrev4-2.bst 2019-01-14 (MD) hand-edited version of apsrev4-1.bst
%Control: key (0)
%Control: author (8) initials jnrlst
%Control: editor formatted (1) identically to author
%Control: production of article title (0) allowed
%Control: page (0) single
%Control: year (1) truncated
%Control: production of eprint (0) enabled
%

\end{document}